\documentclass[twocolumn,aps,prl,preprintnumbers,amsmath,amssymb,showpacs,showkeys]{revtex4}
\usepackage{graphicx}
\usepackage{dcolumn}
\usepackage{bm}
\usepackage{epsfig,wrapfig}
\usepackage{epstopdf} 
\usepackage{fancyhdr}
\usepackage{indentfirst}
\usepackage{titlesec}
\usepackage{amsmath}
\begin{document}
\title{Sound Focusing by Acoustic Luneburg Lens}
\author{Sang-Hoon Kim}
\email{shkim@mmu.ac.kr }
\affiliation{
Division of Marine Engineering, Mokpo National Maritime University, Mokpo 530-729, R. O. Korea
}
\date{\today}
\begin{abstract}
Luneburg lens is a gradient index lens that focuses the incoming wave on the opposite side of the lens without aberration.
 We developed a two-dimensional acoustic Luneburg lens by changing the refractive index of the medium in the lens.
It has a cylindrical shape and is composed of 701 aluminum columns with various radii of less than 1cm.
It focuses the incoming sound wave on the edge of the opposite side of the lens without aberration as well in the frequency range of 1,000 - 3,200Hz.
It increases the amplitude of the incoming sound wave by 3 - 4 times,
and the amplification corresponds to the sound level difference of 10 - 15dB.
The ability of the acoustic Luneburg lens as sonar was checked by VU meters in the air.
Acoustic Luneburg lens has a simple structure and easy to build.
It could be a strong candidate of a next generation of  sonar.
\end{abstract}
\maketitle

Luneburg lens is a GRIN (GRadient INdex) lens and focuses the incoming wave on the opposite side of the lens without aberration.
It was suggested  by Luneburg in the 1940's \cite{lu} and studied by Gutman \cite{gu}
in the 1950's as a part of geometrical optics.
It is a typical model of transformation wave theory.
The general solution of Luneburg lens was studied by Morgan \cite{mo}.
The basic methodology of the lens is as follows:
One conjugate sphere is assumed to be outside the lens, while the other maybe either inside, outside, or at the surface.
If one of the spheres is of infinite radius, then the lens will focus a parallel beam perfectly at a point on the other sphere.
The focusing ability of electromagnetic Luneburg lens has been applied to
communications and nuclear scattering \cite{kim,michel}.

Recently, there has been a lot of progress thanks to metamaterials and transformation optics.
A two-dimensional (2D) broadband planar Luneburg lens based on I-shaped metamaterials was studied by Cheng et al. \cite{ch}.
A Luneburg lens with a zero F-number was studied by Schurig \cite{sc}.
Paths of oblique rays from a distant source are focused onto the flat image plane.
Later, an external-angle broadband Luneburg lens was introduced by Kundtz et al. \cite{ku}.
They showed that even a flattened lens works as well as a Luneburg lens by using a powerful emerging technique.
A couple of the impressive developments of Luneburg lens research were undertaken in electromagnetic waves, too.
Falco et al. fabricated the lens in silicon photonics \cite{fa},
 and Zentgraf et al. in plasmonics in the order of $\mu m$ \cite{ze}.

It is not surprising to try to make an acoustic Luneburg lens (ALL) because it is a direct
acoustic application of transformation optics.
One serious application area of ALL is sonar, because it gives the direction of the acoustic source directly.
It works passively by focusing a sound wave, and actively by creating a plane wave at both modes.
A pioneering study of ALL began long time ago by Boyles \cite{bo}
just after the study of optical Luneburg lens in 1960's.

For acoustic GRIN lens techniques, there have been several research results very recently.
Sound focusing by acoustic GRIN lens was realized by a few groups \cite{cli2010,cherry}.
A 2D cylindrical omnidirectional acoustic absorber, acoustic black hole,
was realized in the air by Climente et al. \cite{cli2012} and under water by Naify et al. \cite{orris}.
A serious progress in GRIN lenses for flexural waves were reported by Climente et al., too \cite{cli2014}.
They successfully produced numerical simulation of ALL by changing the thickness of the plate through periodic arrangements.
However, till now a realization of ALL, that is, sound focusing on the opposite side of the lens
without aberration in macroscopic scale has not been achieved yet.

Applying the previous acoustic GRIN lens methods,
we developed a 2D ALL by changing the density of space inside the lens
because the density is related with the refractive index of the medium.
ALL focused the incoming acoustic wave on the edge of the opposite side of the lens as well and vice versa.
In this paper we report the design and performance of the 2D ALL as sonar in the air.
A dynamic response depending on the movement of the acoustic source like sonar was demonstrated.


Luneburg lens of circularly symmetric shape has the practical advantage of being much more adaptable to rotation.
The index of refraction of a Luneburg lens is given by a function of the radius
\begin{equation}
n(r)=\sqrt{2-\left(\frac{r}{a} \right)^2},
\label{1}
\end{equation}
where $a$ is the radius of the lens and $0 \leq r \leq a$.
It was derived from Fermat's principle and the calculus of variation \cite{lu,mo}.
2D Luneburg lens is cylindrical and 3D Luneburg lens is spherical,
 but they share the same principle of the refractive index in Eq.~(\ref{1}).
We introduce the 2D one here.

Wave equation of an optical Luneburg lens is governed by  permittivity and permeability,
and ALL, the acoustic counterpart, is governed by density and modulus
or inverse compressibility of the medium.
The background speed of sound wave is $v_o=\sqrt{B/\rho_o}$,
where $B$  and $\rho_o$ are the background bulk modulus and density outside the lens.
The modulus is assumed to be constant inside and outside of the lens,
and then the density of the medium is the key of ALL.
The density is a function of the radius only.
At the same time the refractive index in Eq.~(\ref{1}) can be rewritten in the discrete form
\begin{equation}
n_i = \sqrt{2-\left( \frac{i-0.5}{N}\right)^2},
\label{3}
\end{equation}
where $N$ is the number of concentric layers inside the lens and $i=1, 2, ... N$.
Therefore, we obtain the relation between i-th refractive index and i-th density in the lens.
\begin{equation}
n_i = \frac{v_o}{v_i} = \sqrt{\frac{\rho_i}{\rho_o}}.
\label{4}
\end{equation}

\begin{figure}
\resizebox{!}{0.22\textheight}{\includegraphics{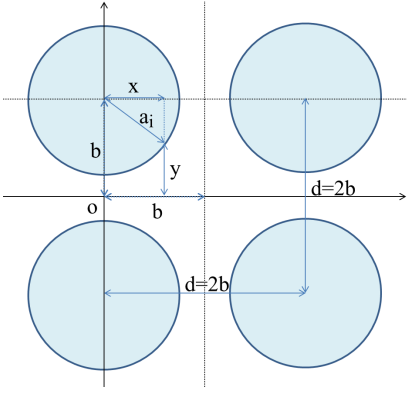}}
\caption{Calculation of the density of the medium using square lattice positioning.}
\label{rms}
\end{figure}

We use a square lattice positioning with circular columns similar with sonic crystals for convenience as in Fig.~\ref{rms}.
Other shape of structure including hexagonal positioning which satisfies the density condition in Eq.~(\ref{4}) will work, too.
The sound wave passes through between the circles, or in fact, the columns of relatively high impedance materials in the air.
The density of the lens depends on the ratio between the distances of the center of the circle and that of the surfaces of the circle as in Fig.~\ref{rms}.
\begin{equation}
\frac{\rho_i}{\rho_o}=\frac{b}{y_{rms}}.
\label{9}
\end{equation}
The distance $y$  is a function of  $x$ in Fig.~\ref{rms}.
 \begin{eqnarray}
y(x) = b & -\sqrt{a^2-x^2}  & :   0 \le x \le a,
\\ \nonumber
     = b &                & :   a \le x \le b.
\label{10}
\end{eqnarray}
The $a$ is the radius of the columns and a function of the index $i$, that is, $a_i$.
Then,  $y_{rms}$ in Eq.~(\ref{9}) is obtained by a simple integration.
\begin{eqnarray}
y_{rms}^2  &=& \frac{1}{b} \left\{ \int_0^a (b-\sqrt{a^2 - x^2})^2 dx + \int_a^b b^2 dx \right\}
 \\
\label{11}
     &=&  b^2 -\frac{\pi a^2}{2}   + \frac{2 }{3 } \frac{a^3}{b} .
\label{12}
\end{eqnarray}

From  Eqs.~(\ref{3}) and (\ref{12}), we obtain the $n_i^2$.
\begin{equation}
2-\left( \frac{i-0.5}{N}\right)^2
= \frac{1}{\sqrt{1-\frac{\pi}{2}w_i^2 + \frac{2}{3}w_i^3}},
\label{16}
\end{equation}
where $w_i =a_i/b$. Note that $0 < w_i < 1$.
The ratio $w_i$ is obtained by solving the nonlinear equation of Eq.~(\ref{16}).
We set b=1cm and the number of concentric rings as $N=15$.
Then, distance between columns as d=2b=2cm and the radius of the lens is R=Nd=30cm.

\begin{figure}
\resizebox{!}{0.28\textheight}{\includegraphics{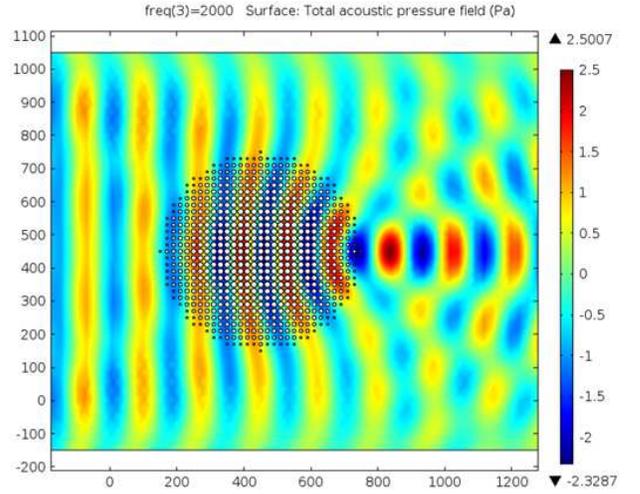}}
\caption{Numerical simulation of the acoustic Luneburg lens
 which was composed of 701 aluminum columns with different radii.
 f = 2,000Hz, The unit is mm. }
\label{simul2}
\end{figure}

Eventually, we obtained a numerical simulation of the ALL at some frequency ranges in Fig.~\ref{simul2}.
Previous numerical simulation was a direct substitution of the refractive index
in Eq.~(\ref{1}) into a computer simulator \cite{cli2014}.
However, that does not give how to control the refractive index in macroscopic scale realization.
The structure can be considered as a homogenized medium made of subwavelength discrete units.
The available frequency range for focusing is decided by the structure of the lens as
4d $ < \lambda < $2R for very long cylinder of the height $h>D$.
The 4d comes from the homogeneity condition of wave.

\begin{figure}
\resizebox{!}{0.2\textheight}{\includegraphics{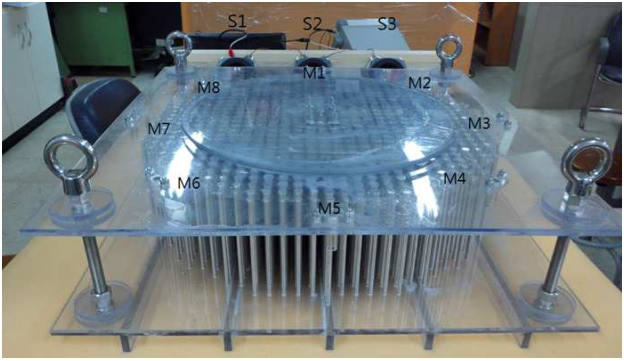}}
\caption{Experimental setup of the cylindrical acoustic Luneburg lens.
Three speakers and eight microphones were installed in equidistant intervals.
}
\label{photo}
\end{figure}

Following the numerical simulation, we made ALL using 701 aluminum columns of various radii of less than 1cm in Fig.~\ref{photo}.
Besides aluminum, any material which has large acoustic impedance will work.
The height of the cylindrical lens was 15cm and the bottom
and roof were restricted by two light-transparent FRP (Fiber Reinforced Plastics) panels.
Three speakers of S1, S2, and S3 with $4\Omega - 25W$ were used to create pure tone plane waves
and eight microphones from M1 to M8 were installed and connected to an oscilloscope as the receivers.
The distance between S2 and M1 was 30cm.

\begin{figure}[h]
  \centering
\resizebox{!}{0.19\textheight}{\includegraphics{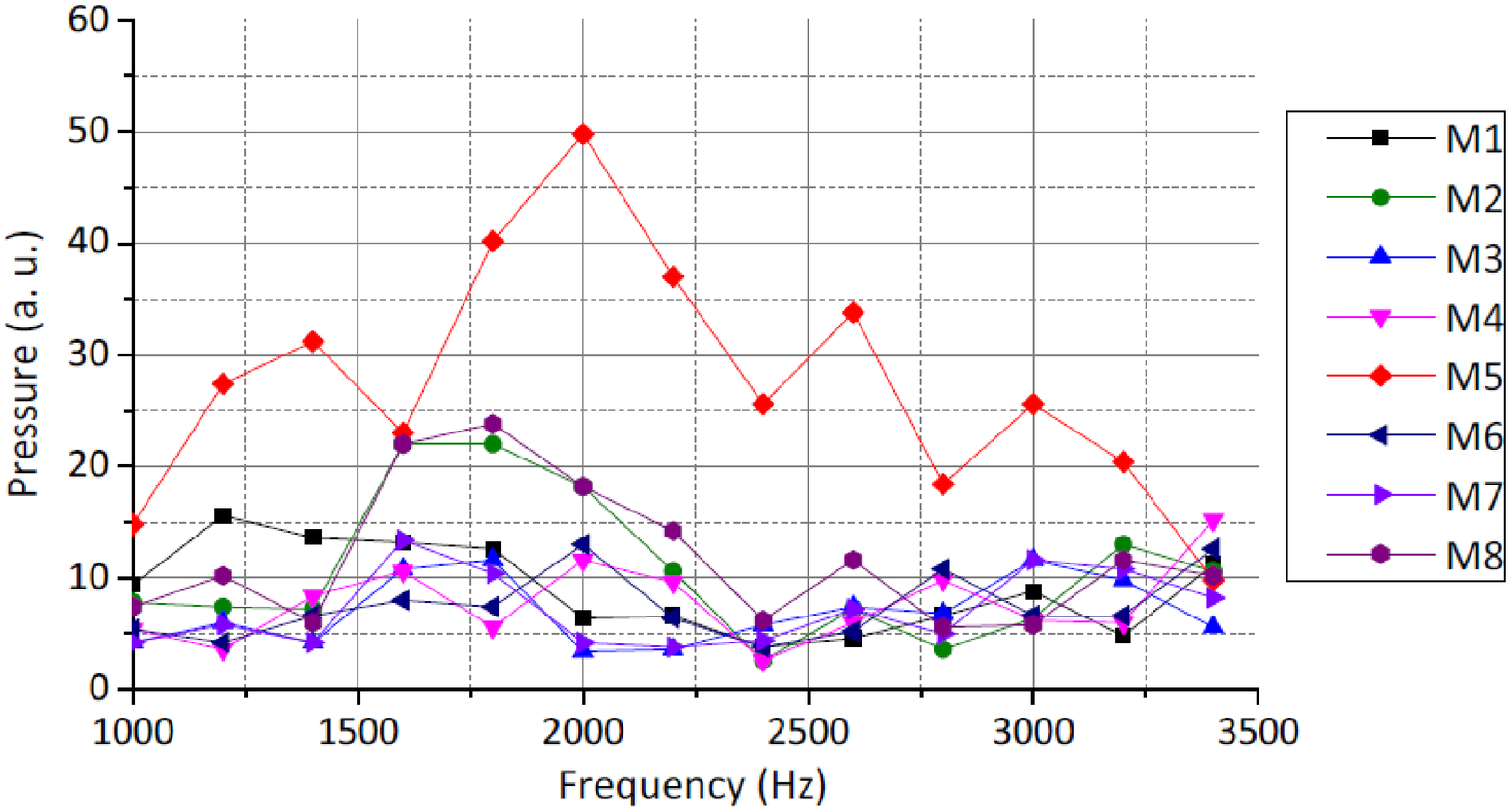}}
  \caption{Acoustic pressures or output voltages at eight positions on the edge. }
  \label{pressure}
\end{figure}

\begin{figure}
\resizebox{!}{0.20\textheight}{\includegraphics{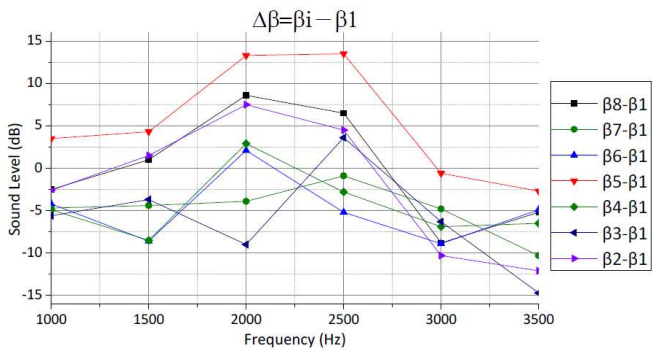}}
\caption{Sound level differences compared with the nearest receiver M1.
 }
\label{soundlevel}
\end{figure}

The acoustic pressures of the eight positions were measured as voltages of $V_{pp}$
with $\Delta f = 200Hz$ steps in Fig.~\ref{pressure}.
It depends on the input voltage and sensitivity of the speakers.
The ALL increases the acoustic pressure of the incoming wave by 3 - 4 times.
The ALL has 90 degree rotation symmetry, but the rotation of 45 degree did not make a big difference.
We switched the position of emitters and receivers, and found that the ALL produces plane waves outside the lens perfectly.
It means that if we use ALL as sonar, it can be used as passive and active modes both.

We measured the sound level by using sound level meter independently in Fig~\ref{soundlevel}.
It increased the sound level by about 10 - 15dB at the focusing point M5 compared with the nearest point M1.
It  matches the relation $\Delta \beta = 20 \log (V_{pp}(i)/V_{pp}(1))$ approximately.
The effective distance between the speakers and the ALL was restricted by the two rigid FRP panels
and the sensitivity of the microphones.

\begin{figure}
\resizebox{!}{0.20\textheight}{\includegraphics{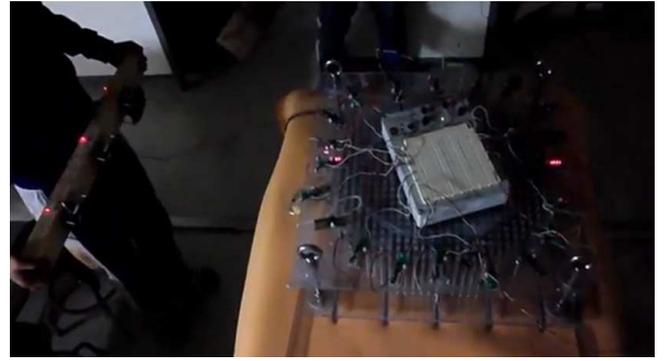}}
\caption{Demonstration of Luneburg sonar.
See supplementary material for full movie \cite{tube}.}
\label{movie}
\end{figure}

The film of the ALL's performance as sonar using VU meters or sound-to-light converters
was made \cite{tube}.
The motion of the acoustic source is observed directly by naked eyes.
As approaching the source, the VU meters are the brighter, and as receding the source, the VU meters are dimmer.
The distance between the sound source and the lens is so close that the effect of the distance is significant in the film.
If the acoustic source is far away, then the output voltage of M5 in Fig.~\ref{photo} will be much dominant.


We constructed a 2D  ALL by variable density method inside the lens.
It is composed of 701 aluminum columns with various radii of less than 1cm
on the square lattice positioning.
It has concentric 15 layers and the diameter of the lens is 60cm.
It increases the amplitude of the incoming sound wave by 3 - 4 times,
and the amplification corresponds to the sound level difference of 10 - 15dB.
It works as sonar by focusing the incoming wave on the edge of the opposite side of the lens without aberration as well in the frequency range of 1,000 - 3,200Hz.
The performance was checked by sound-to-light converters.
The working distance between emitters and receivers was increased by extending the panels.

We tested the performance of the ALL as sonar in the air, but it should work underwater as well \cite{orris} and it may be used as a next generation of sonar.
The application area of ALL is not restricted to sonar.
It could be an acoustic window of an underwater which has little or no light, too.
The amplifying ability of ALL gives a possibility of hearing aid operating without battery.
The sound focusing ability can be applied to
an acoustic generator \cite{wang}, street lamp operated by road noise, audio phone without power source, and a long-range super ear, etc.

\acknowledgements
The author acknowledges Jose S$\acute{a}$nchez-Dehesa for comments and discussions.
This research was supported by Basic Science Research Program
through the National Research Foundation of Korea (NRF)
funded by the Ministry of Education (2014-055046).
 
\end{document}